# A New Framework for Understanding Systematic Errors in Cluster Lens Modeling. IV. Constraints from the Giant Arc in Abell 370

Lana Eid and Charles R. Keeton
Rutgers University–New Brunswick, Department of Physics and Astronomy, 136 Frelinghuysen Road, Piscataway, NJ 08854, USA

## Abstract

We aim to improve cluster lens modeling and source reconstruction by utilizing the full information in giant, caustic-crossing arcs lensed by galaxy clusters. Lens models are generally constrained using image positions and assuming point sources, but spatially extended giant arcs provide more constraints; however, they require a more complex model that accounts for the structure of the extended source. We seek to determine whether improvements to the lens model and reconstructed source merit the difficulty of handling the extra constraints. We choose the spatially extended $z = 0.725$ giant arc in the $z = 0.375$ Abell 370 galaxy cluster field for our study. We present (1) a series of pixel-based source reconstructions (PBSRs) for cluster mass models exploring the range of uncertainties in our fiducial model, (2) a similar analysis done using a prototype Python de-lensing code for cluster mass models from each of the Hubble Frontier Fields (HFF) modeling teams, (3) an optimized model with PBSR, and (4) and an investigation of how our optimized model affects the cluster mass model locally and globally in the highest-magnification regions. We find that our optimized model (1) is able to correct resolution-limited assumptions in cluster model inputs local to the arc, (2) has significantly smaller arc model residuals than results from the standard HFF models, and (3) affects the critical curves and therefore the information derived from highest-magnification zones most significantly in regions local to the arc.

*Unified Astronomy Thesaurus concepts:* Gravitational lensing (670); Strong gravitational lensing (1643); Abell clusters (9); Luminous arcs (943)

## 1. Introduction

Gravitational lensing by galaxy clusters magnifies and distorts distant objects into one or more images (R. Narayan & M. Bartelmann 1996; J. Wambsganss 1998; J.-P. Kneib & P. Natarajan 2011; S. Vegetti et al. 2024), providing (1) a way of inferring the underlying dark matter distribution of the lensing cluster (e.g., D. Coe et al. 2010; C. A. Raney et al. 2020a; J. M. Diego et al. 2023b; S. Vegetti et al. 2024; B. Beauchesne et al. 2024; D. Perera et al. 2024) and (2) a uniquely higher-resolution view of high-redshift galaxies than otherwise possible (e.g., J. Richard et al. 2010; J. M. Diego et al. 2016, 2023a; T. L. Johnson et al. 2017a, 2017b; V. Patrício et al. 2018; P. S. Kamieneski et al. 2023; B. Welch et al. 2023).

We can use strong lensing to study the nature (e.g., T. Broadhurst et al. 2025), distribution (e.g., S. Vegetti & L. V. E. Koopmans 2009; L. Dai et al. 2018; L. Ji & L. Dai 2025; T. Broadhurst et al. 2025; M. Limousin et al. 2025), and kinematics (e.g., B. Beauchesne et al. 2024) of dark matter in galaxies and galaxy clusters on the continuous range of scales that is predicted by hierarchical structure formation in cold dark matter (CDM) models (W. H. Press & P. Schechter 1974), especially in and around critical curves in the image plane. These high-magnification regions are of further interest because of the high-redshift objects they magnify. In a cosmological context, studying redshifts $z \sim 6-11$ is key to constraining the timescale and end of the epoch of reionization (H. Atek et al. 2015; J. M. Lotz et al. 2017). Lensing affords us the ability to study galaxies at the faint end of the luminosity function at these redshifts, galaxies that are progenitors of what may become Milky Way–luminosity ($\sim L^*$) galaxies today (J. M. Lotz et al. 2017; H. Atek et al. 2018). It also leads us to various applications such as searching for ancient stars (e.g., J. M. Diego et al. 2023a), predicting new appearances of magnified high-$z$ supernovae (e.g., J. M. Diego et al. 2016; S. A. Rodney et al. 2021; J. D. R. Pierel et al. 2024), resolving the smallest scales of star formation (B. G. Elmegreen & D. M. Elmegreen 2005) in the early Universe (e.g., T. L. Johnson et al. 2017b; Y. Fudamoto et al. 2025), and molecular gas and star formation in the context of the Kennicutt–Schmidt relation (M. Schmidt 1959; R. C. Kennicutt 1989, 1998; C. E. Sharon et al. 2019).

Galaxy cluster lens models constructed from data are required to interpret the cluster mass distribution as well as accurately de-lens the observed images. Cluster lens models are generally constrained using the positions and relative brightness of point-like image centers and identifiable clumps from strongly lensed background sources (e.g., J. Richard et al. 2010; T. L. Johnson et al. 2017a, 2017b; S. J. Dunham et al. 2019; C. A. Raney et al. 2020a; P. Bergamini et al. 2023), ideally with spectroscopic redshifts; statistical analysis of the target field can offer weak lensing data for additional constraints. Galaxy cluster lenses have a more complicated mass distribution than smaller-scale galaxy lenses, and with that complexity comes a much larger number of estimated and free parameters to fit.

Given the extent and quality of new, high-resolution data from programs such as the Hubble Frontier Fields (HFF; J. M. Lotz et al. 2017), C. A. Raney et al. (2020b) found that statistical uncertainties have become smaller than the remaining systematic differences between different cluster modeling techniques (e.g., parametric, nonparametric, hybrid, etc.). This implies imperfect knowledge about the mass distribution of the cluster, whose larger-scale mass components are dominated by dark matter.

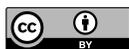

However, there is information in the rest of the structure of giant arcs from extended sources that can be used to better constrain the lens model. Caustic-crossing sources are lensed into giant arcs that cross image plane critical curves, regions representing theoretically infinite magnification specific to each pair of cluster–source redshifts. Conclusions drawn about lensed objects projected on and near these critical curves are sensitive to the calculated cluster mass model. For example, Y. Fudamoto et al. (2025) noted that none of the available cluster mass models for the Abell 370 (A370) galaxy cluster were able to explain some of the microlensed stars in the giant arc that they observed with JWST; events close to the critical curves were consistent with the mass models, but farther away microlensed stars could be (1) interpreted as evidence of different dark matter structure, substructure, or distribution than originally assumed (J. M. Diego et al. 2024; M. Limousin et al. 2025), (2) interpreted as testable evidence of non-CDM (T. Broadhurst et al. 2025), (3) interpreted as evidence of a shallow stellar luminosity function (J. M. Diego et al. 2024), or (4) misinterpreted in the case that the mass model is incorrect. Developing a method that is able to utilize the full arc information in the process of constraining the mass model can facilitate accurate interpretations of such events.

We seek to determine whether the difficulty of handling the complex parameters in an extended source model would improve both the cluster mass model and the reconstructed source galaxy. It has historically been more computationally expensive to include all the pixels of extended brightness distributions of lensed objects as additional constraints (i.e., A. S. Tagore & C. R. Keeton 2014; J. M. Diego et al. 2016). In addition, utilizing source reconstruction in fitting introduces more systematic effects from choices such as gridding and regularization techniques (e.g., A. S. Tagore & C. R. Keeton 2014; L. Yang et al. 2020; A. Galan et al. 2024) that attempt to correct but not overfit things such as noise in the data. Using the pixel-based source reconstruction (PBSR) code `pixsrc` (A. S. Tagore 2014; A. S. Tagore & C. R. Keeton 2014) as a starting point, our demonstrative study utilizes a method that would primarily help improve constraints on the regions of higher magnification probed by the specific cluster–background source redshift pairs; we investigate the new changes with respect to (1) the assumptions made in the priors and inputs to the mass model and (2) any scientific conclusions about the lens or the source that could be calculated using these results.

Some examples of using extended source models to constrain cluster lens models can be found in the literature, though their methodology is sufficiently different to not preclude our extended source analysis. (1) J. M. Diego et al. (2016) was able to correct unphysical predictions for the time-delayed reappearance of the first observed multiply lensed supernova Refsdal in a resolved galaxy lensed by the MACS J1149.5+2223 galaxy cluster. They used matching star-forming regions in each image along with an initial guess for the size and shape of their extended source, treating it as a collection of point sources in a Navarro–Frenk–White profile, to improve their nonparametric lens model and resulting source reconstruction; they were able to predict when and where the supernova would appear in the other images with up to about 7% accuracy (J. M. Diego et al. 2016). (2) L. Yang et al. (2020) used a forward-modeling source reconstruction code `LENSTRUCTION`, which works with the lens modeling code `LENSTRONOMY` (S. Birrer et al. 2015; S. Birrer & A. Amara 2018) in an effort to reduce these differences and have different cluster models converge on similar solutions, and they had more success using the extended and resolved lensed images. However, their code only handled separate, multiply imaged sources; our more general method handles the more complex, caustic-crossing background sources that clusters lens into giant arcs, probing some of the higher-magnification regions that C. A. Raney et al. (2020b) found had the most uncertainties between the models. (3) A. Galan et al. (2024) performed a source reconstruction on the arc termed "El Anzuelo" ("The Fishhook") in the galaxy cluster "El Gordo" from JWST data to further constrain their cluster lens model and study the source properties; however, "El Anzuelo" is an arc that is primarily lensed by a few local galaxies. The arc we will be using in our study is one that is clearly lensed by both the global dark matter halos in the cluster as well as the cluster galaxies local to it, so it can inform us about which part of the lens model may be incorrect if available giant arcs and extended lensed images are not taken into account.

We take advantage of the high-resolution HFF (J. M. Lotz et al. 2017) observations of the A370 galaxy cluster ($z = 0.375$), which contains one of the largest and first known giant arcs ($z = 0.725$) in a cluster field (G. Soucail 1987; G. Soucail et al. 1987, 1988). This giant arc, shown in Figure 1 and sometimes termed "The Dragon," is comprised of five merging images of the same background galaxy. The leftmost portion of the arc shows a complete image of the background source, while the most magnified parts of the image (up to magnifications of about 30) are along the thinner, more stretched section of the arc (J. Richard et al. 2010; V. Patrício et al. 2018).

Two previous source reconstructions of the giant arc in the A370 field both used the `lenstool` code (E. Jullo et al. 2007; J.-P. Kneib & P. Natarajan 2011) on HST data, with the first lens model for the cluster by J. Richard et al. (2010) being updated by V. Patrício et al. (2018); both mainly used the leftmost image in their source reconstructions. J. Richard et al. (2010) used the center positions of each image as well as the locations of three bright, resolved star-forming clumps that each appear in three of the five images as additional constraints. V. Patrício et al. (2018) made use of MUSE [O II] data to update their mass model and reconstruct the source in different filters; they used this to study the star formation causing the [O II] lines in the source galaxy, calculated extinction and star formation rate density maps for the source, and constructed velocity maps from the [O II] data.

Our goal is to determine whether the information of the extended giant arc can be used to improve constraints on the lens model for A370. In order to do this, we perform a series of full PBSR of the source of the giant arc in A370 using `pixsrc` (A. S. Tagore & C. R. Keeton 2014), which works in conjunction with `lensmodel` (C. R. Keeton 2001, 2011). The `pixsrc` code reconstructs the source on a grid and reproduces the data pixel by pixel; it ray-traces each image pixel back to the source plane and uses the determined magnification there to adjust the resolution of the grids used to reconstruct the source. We also introduce a new Python de-lensing code for use with lens model results provided by other modeling teams. Section 2 reviews the key methods for lens modeling and PBSR, while Section 3 provides background on the HFF data and lens models that are used in our analysis. Section 4 presents our results, and Section 5 summarizes our findings. We assume a $\Lambda$CDM cosmological model and use approximate values from Planck Collaboration et al. (2016) for the composition of the Universe: $\Omega_m = 0.3$, $\Omega_\Lambda = 0.7$, and $h = 0.7$.

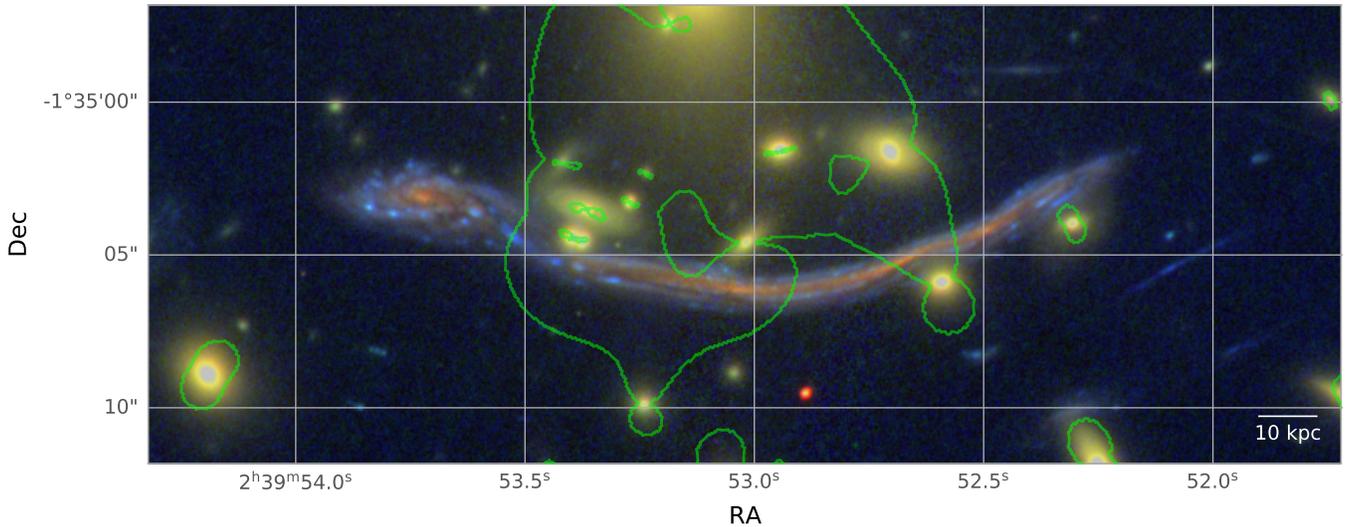

**Figure 1.** Close-up of the lensed arc in a red–green–blue (RGB) image using Hubble Space Telescope (HST) F814W, F606W, and F435W filters. The full cluster field can be seen in Figure 2. Green curves show the lensing critical curves for the cluster–source redshift pair using our fiducial lens model from C. A. Raney et al. (2020a). A 10 kpc scale bar is shown for reference.

## 2. Methods

### 2.1. Key Lens Theory

The lens equation is the foundation of gravitational lensing:

$$\boldsymbol{\beta} = \boldsymbol{\theta} - \frac{D_{LS}}{D_S} \boldsymbol{\alpha}. \quad (1)$$

Here $\boldsymbol{\theta}$ and $\boldsymbol{\beta}$ are the angular positions in the image and source planes, respectively, and $\boldsymbol{\alpha}$ is the lensing deflection. In this formulation, the deflection is scaled to a source at infinity; it can be rescaled to any finite redshift through the distance ratio $D_{LS}/D_S$ as shown, where $D_S$ and $D_{LS}$ are, respectively, the angular diameter distances from the observer to the source and from the lens to the source. The deflection can be written in terms of the two-dimensional effective lensing potential $\phi$ as

$$\boldsymbol{\alpha} = \boldsymbol{\nabla}_\theta \phi. \quad (2)$$

The HFF lens model outputs include maps for the two components of deflection ($\alpha_x$ and $\alpha_y$) as well as the lensing convergence ($\kappa$) and the two components of shear ($\gamma_1$ and $\gamma_2$). The latter three quantities can be written in terms of second derivatives of the lensing potential as

$$\kappa = \frac{\phi_{11} + \phi_{22}}{2}, \quad (3)$$

$$\gamma_1 = \frac{\phi_{11} - \phi_{22}}{2}, \quad (4)$$

$$\gamma_2 = \phi_{12} = \phi_{21}. \quad (5)$$

Once the convergence and shear have been scaled to the lens–source redshift pair, the lensing magnification can be computed as

$$\mu = \frac{1}{(1-\kappa)^2 - \gamma_1^2 - \gamma_2^2}. \quad (6)$$

Lensing critical curves are locations in the image plane where the magnification diverges, so $\mu^{-1} = 0$. The critical curves map to caustics in the source plane.

A key feature of gravitational lensing is conservation of surface brightness. If we take $S(u, v)$ to be the surface brightness distribution in the source plane, we can compute the corresponding brightness in the image plane as

$$I(x, y) = S(x - \alpha_x, y - \alpha_y). \quad (7)$$

### 2.2. Lens Models

We primarily use HFF lens modeling results for A370 from C. A. Raney et al. (2020a, 2020b), who make use of the fully parametric `lensmodel` code (C. R. Keeton 2001, 2011). Each mass model includes 275 total mass components: four mass components representing the complicated cluster dark matter halo, 256 cluster member galaxies, and 15 line-of-sight galaxies that C. A. Raney et al. (2020a) identified as being important for the lens model. The masses and truncation radii of the galaxies are calibrated using mass–luminosity scaling relations following F. Brimioulle et al. (2013). We use the `pixsrc` code from A. S. Tagore & C. R. Keeton (2014) in conjunction with the cluster lens model to perform a series of PBSR and lens model improvements.

Other HFF lens modeling teams adopted their own approach to parametric lens models. In addition, several teams used nonparametric, or free-form, models in which the model is described in terms of mass pixels or other basis functions. Hybrid methods mix parametric and nonparametric

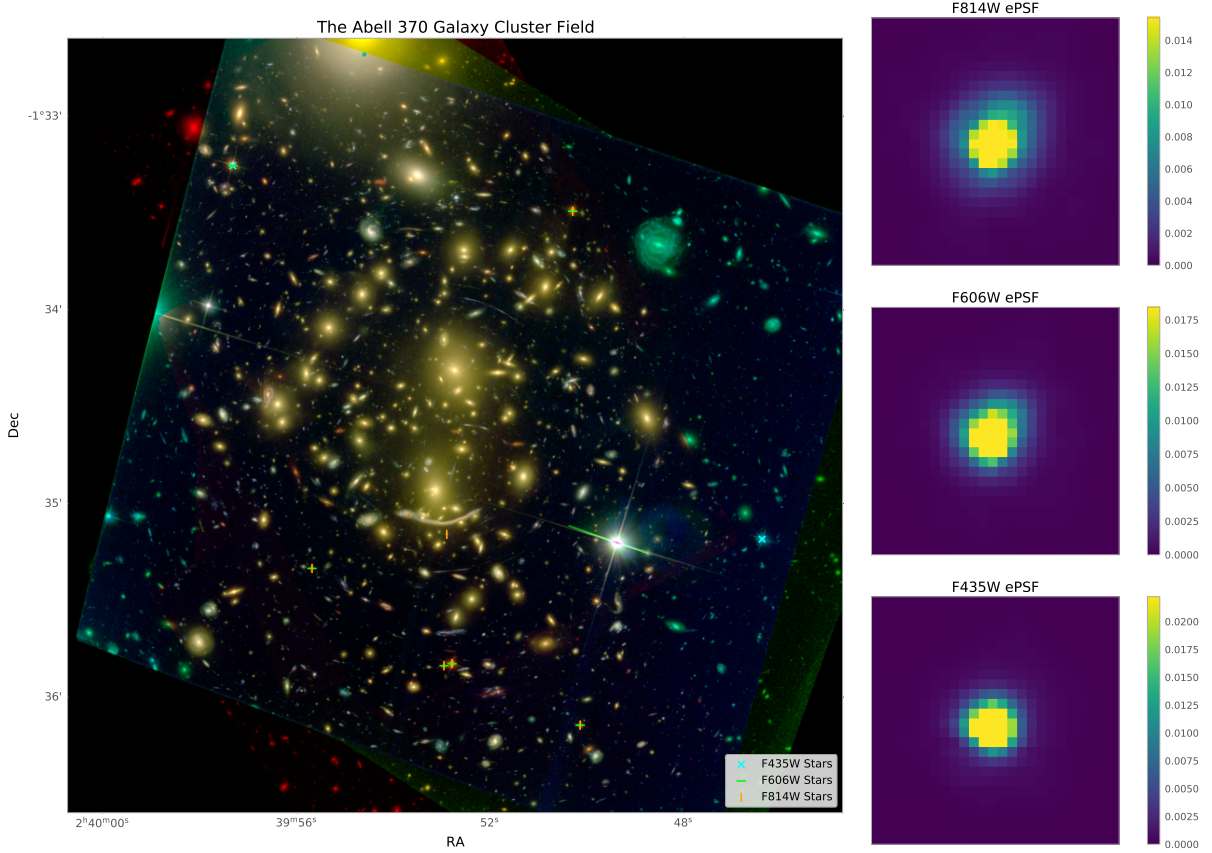

**Figure 2.** Left: the assembled RGB image from Trilogy. Stars in the F814W filter are marked by orange vertical lines, stars in F606W are marked by green horizontal lines, and stars in F435W are marked by blue crosses. Right: empirical PSFs for the three filters obtained using `EPSFBuilder`; each panel is about $0\rlap{.}''76$ on each side.

techniques. The specific set of HFF lens models that we used for A370 are listed in Section 3.1.

### 2.3. Pixel-based Source Reconstruction

Here we review key concepts for PBSR. The full formalism can be found in S. H. Suyu et al. (2006) and A. S. Tagore & C. R. Keeton (2014), and more details can be found in the literature (S. Wallington et al. 1996; C. R. Keeton 2001; S. J. Warren & S. Dye 2003; S. Dye & S. J. Warren 2005; L. V. E. Koopmans 2005; B. J. Brewer & G. F. Lewis 2006; S. Vegetti & L. V. E. Koopmans 2009; A. S. Tagore & N. Jackson 2016).

Let $s$ be a collection of pixel surface brightness values in the source plane, and $d$ be the corresponding pixel surface brightness values in the image plane. Because lensing conserves surface brightness, the two quantities can be related to one another by a linear equation of the form

$$d = s + Ln, \quad (8)$$

where $L$ is the "lensing operator" for a given mass distribution and $n$ denotes the noise (S. H. Suyu et al. 2006). The lensing operator can include blurring by a point-spread function (PSF):

$$L = L_{\text{PSF}} L_{\text{lens}}. \quad (9)$$

It is conventional to define a penalty function as follows:

$$\chi^2 = (Ls - d)^{\mathrm{T}} C_D^{-1} (Ls - d) + \lambda s^{\mathrm{T}} H^{\mathrm{T}} H s. \quad (10)$$

The first term quantifies how well the model image $Ls$ matches the data $d$ given the noise covariance matrix $C_D$. The second term represents "regularization," which is meant to avoid overfitting noise by penalizing deviations from a smooth source. The $H$ operator can be chosen to regularize based on the surface brightness distribution itself or its first or second derivative (see, e.g., S. J. Warren & S. Dye 2003; S. H. Suyu et al. 2006). The coefficient $\lambda$ is known as the "regularization strength." We can find the optimal source by solving

$$\frac{\partial \chi^2}{\partial s} = 0. \quad (11)$$

Since $\chi^2$ is quadratic in $s$, this amounts to solving the linear equation

$$(L^{\mathrm{T}} C_D^{-1} L + \lambda H^{\mathrm{T}} H) s = L^{\mathrm{T}} C_D^{-1} d. \quad (12)$$

If we choose a lens model, which determines $L$, and a value for the regularization strength $\lambda$, we can solve for the source $s$. We next vary $\lambda$, with $L$ fixed, to maximize the Bayesian evidence. Then, we can repeat the process across the lens model parameter space to find the overall best fit. We use the implementation of this in the `pixsrc` code (A. S. Tagore & C. R. Keeton 2014; A. S. Tagore & N. Jackson 2016), wherein we specifically use the second-order regularization mode that penalizes large values of the Laplacian of the source surface brightness distribution.

### 2.4. Python De-lensing Code Prototype

In order to use `pixsrc`, we need a complete specification of the lens model. We have that information for the HFF lens models from our team, but not for the other teams. We therefore need an alternate, but hopefully equivalent, method

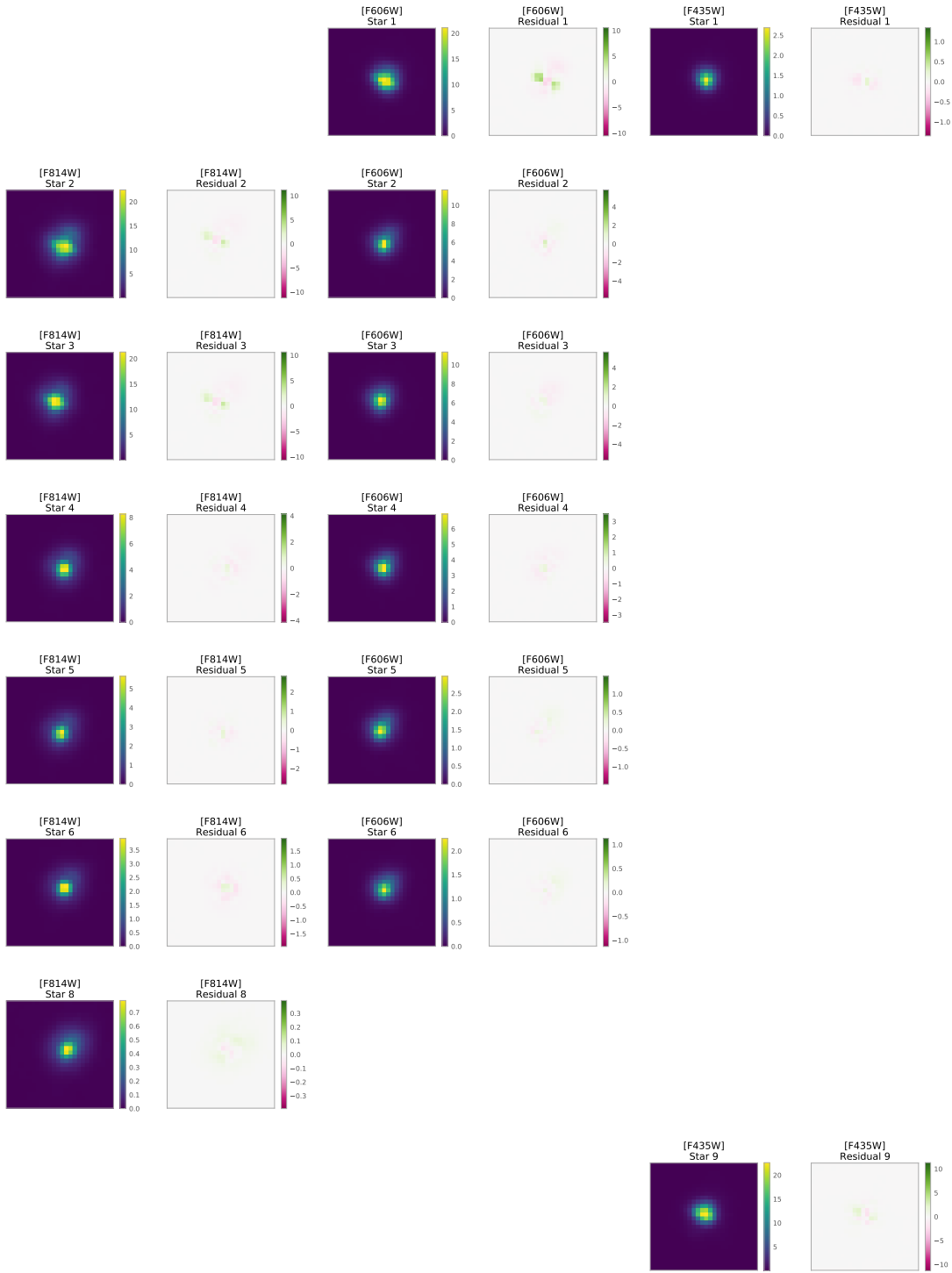

**Figure 3.** Validation of the ePSFs. Each row depicts a different star identified in Figure 2. Columns 1, 3, and 5 show individual star cutouts in the different filters, while columns 2, 4, and 6 show residuals after subtracting a scaled ePSF.

to work with lens models for which we have maps of deflection, convergence, and shear.

Fortunately, it is straightforward to build a simplified source reconstruction code if blurring from atmospheric and optical effects is mild. We have built an initial version of this code in Python that includes the following steps.

1. Deconvolve the image from the PSF using Richardson–Lucy deconvolution with the `skimage.restoration.richardson_lucy` routine.

2. De-lens the image by mapping all of the image plane pixels back to the source plane. The resulting set of points will have irregular spacing, so we use `scipy.interpolate.griddata` to interpolate onto a uniform grid in the source plane. The interpolation serves as a form of regularization.

3. Relens the interpolated source and convolve with the PSF to obtain the model image.

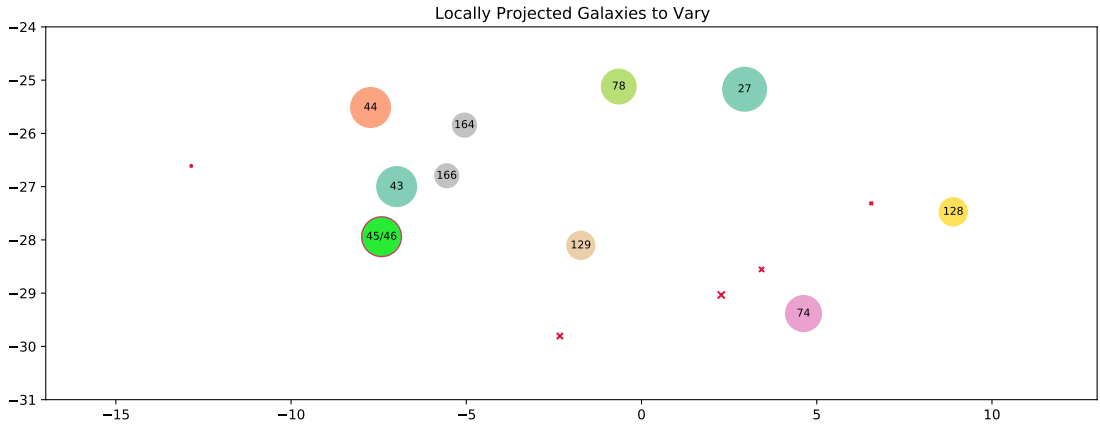

**Figure 4.** Locations of the galaxies we modeled and subtracted (the BCG we subtracted is not visible in this close-up image). These are also the galaxies that are allowed to vary in the revised lens models (see Section 4.3). Each colored circle corresponds to a different cluster member in our fiducial lens model, with the size of the circle corresponding to the lensing strength assigned. Note that the circle labeled "45/46" has traditionally been treated as a single mass in lens models but is actually resolved into two galaxies. The five nominal image centers for the arc are indicated with red × marks. The axes are in arcseconds with respect to the reference coordinates for the field.

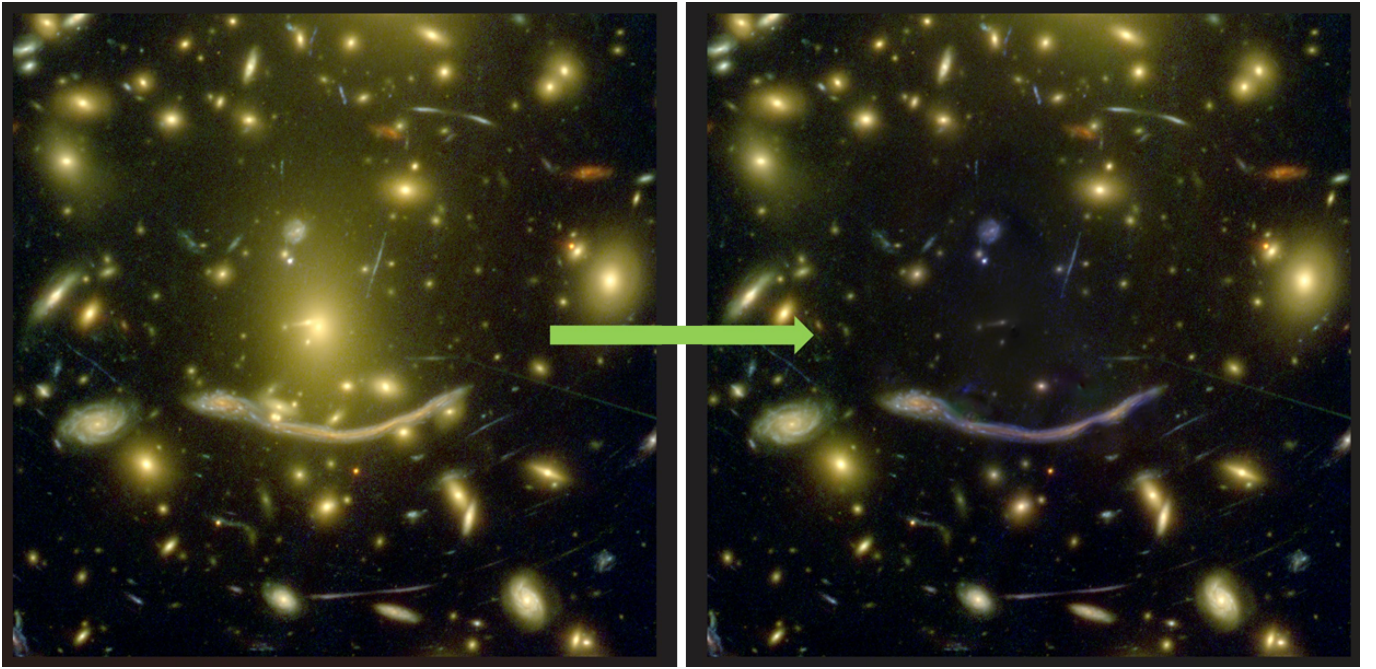

**Figure 5.** Left: original image of the arc shown for part of the A370 galaxy cluster field from the HFF. This RGB image was constructed using data from HST filters F435W, F606W, and F814W. Right: cleaned image after subtracting 12 elliptical cluster members contaminating the arc light, including one of the two BCGs. Note that the two small galaxies near the bend in the arc, shown as removed here, were not modeled and removed in the actual source reconstructions, as we found that excluding them with our data mask (Figure 6) produced better results.

This analysis can work entirely on maps, namely maps of the data and PSF, as well as maps of the two components of deflection; it can also be used with input shear and convergence maps to plot critical curves and caustics on the image plane and source plane, respectively. We test this new code against `pixsrc` in Section 4.1 below.

## 3. Data

### 3.1. Hubble Frontier Fields Data and Lens Models

The HFF program was a 3 yr, 840 orbit program that used the HST and the Spitzer Space Telescope to obtain deep images of six galaxy clusters and the faint $z \sim 5-10$ background galaxies that have been magnified by lensing (J. M. Lotz et al. 2017). This program included a lens modeling effort by a set of teams with different modeling approaches. Each HFF team provided a standard set of outputs for a fiducial lens model as well as a set of "range" models that sampled the uncertainties in their models. The model outputs were calibrated to a source at infinity but can be scaled to any appropriate lens–source redshift combination through the angular diameter distance ratio $D_{LS}/D_S$ (see Section 2.1). All results in this paper are scaled to the A370 arc using the cluster redshift $z_L = 0.375$ and the arc redshift $z_S = 0.725$.

We used publicly available HFF lens model results in the form of deflection, convergence, and shear maps.[1] Five teams used parametric lens models: Caminha (G. B. Caminha et al. 2017), Clusters As Telescopes (E. Jullo et al. 2007; M. Jauzac et al. 2012, 2014; J. Richard et al. 2014), Glafic (M. Oguri 2010; M. Ishigaki et al. 2015; R. Kawamata et al. 2016, 2018), Keeton (C. A. Raney et al. 2020a), and Sharon (E. Jullo et al. 2007; T. L. Johnson et al. 2014). Two teams used nonparametric lens models: Bradač/Strait (M. Bradač et al. 2005, 2009) and Williams (J. Liesenborgs et al. 2007). Finally, the Diego team (J. M. Diego et al. 2005a, 2005b, 2007, 2015; J. Vega-Ferrero et al. 2019) used a hybrid technique.

---
[1] https://archive.stsci.edu/prepds/frontier/lensmodels/

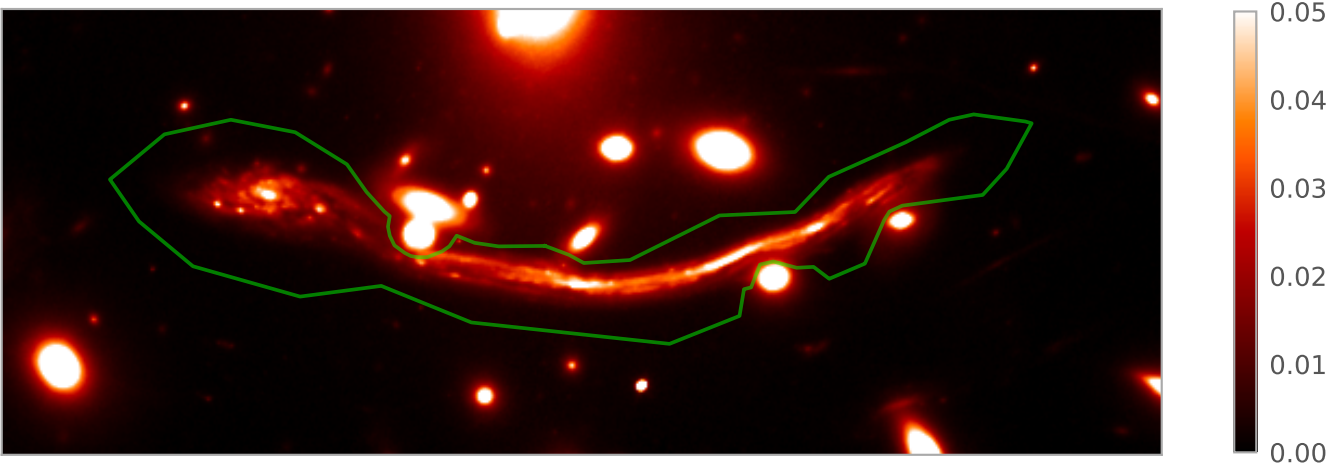

**Figure 6.** Details of the data used to evaluate the lens models. The map shows the data in the F814W filter (prior to the removal of contaminating light), and the green polygon shows the data mask (created after the galaxy modeling and removal).

### 3.2. Empirical Point-spread Function

The `pixsrc` code requires a PSF for each data image it uses. Our RGB image contains exposures from HST filters F814W (∼red), F606W (∼green), and F435W (∼blue). We use the `EPSFBuilder` function from the Python `Photutils` (L. Bradley et al. 2022) package to create an "effective PSF" (ePSF) for each filter. We first choose unsaturated, relatively isolated stars from C. A. Raney et al. (2020a)'s results with which to build this ePSF. This is a combination of objects flagged with high enough "stellarity" in Trilogy[2] and additional stars chosen through visual inspection. Our field is crowded due to it spanning most of a galaxy cluster, so our star sample is somewhat limited. Since each filter covered a slightly different area, and since each individual star is bright in some bands and dim in others, there is some but not much overlap in the manually confirmed "good" stars between the filters, as seen in the first, third, and fifth columns of Figure 3. After background subtraction using sigma clipping, including only median values in each star cutout, and recentering the ePSF, the effective/empirical PSF for each filter that was autonormalized by `EPSFBuilder` can be seen in the right panel of Figure 2.

We test each PSF by producing residual images and comparing them to the individual stars used to fit it; this can be seen in Figure 3 for the F814W filter (first and second columns), F606W filter (third and fourth columns), and F435W filter (fifth and sixth columns). The major features in the residuals are from other stars and objects prior to removal, but the features that are from the centered star shape before the ePSF is scaled and subtracted are much dimmer by comparison to the original star cutout. The star cutouts that are most isolated with no oversaturated pixels perform similarly to the ePSF for that filter, as can be seen by the smaller and less extended residuals in the second, fourth, and sixth columns of Figures 3. We confirm that our empirical PSFs are reasonable for their intended purpose by separately convolving a mock Sérsic galaxy with each ePSF and star cutout and comparing residuals, where we confirm that the star cutouts most similar to the ePSFs in their respective filters lead to minimal residuals.

### 3.3. Galaxy Modeling and Masking

Since we only want light from the giant arc to be traced back in the PBSR process, we need a cleaned, isolated image of our arc. The light contamination in our arc image is mainly due to elliptical cluster members projected near or on the arc. We use the `isophote` package from Python `Photutils` (L. Bradley et al. 2022) to model and remove the contaminating objects. The `Ellipse` class from the `isophote` package fits a series of elliptical isophotes to a galaxy given an initial guess for a single elliptical isophote; it then builds a model of the galaxy corresponding to those isophotes. We then take this model, adjust it for the local background value (which varies slightly throughout the image), and subtract it from the image.

We do this process separately for drizzled HFF images from three HST filters: F814W (red), F606W (green), and F435W (blue). The light from each individual contaminating galaxy has a different spatial extent in each wavelength depending on what process in the galaxy is emitting that light, so the modeling and removal steps must be followed separately for each filter. The A370 galaxy cluster has two brightest cluster galaxies (BCGs), one of which contaminates the arc light. Since its light also affects all the other cluster members we need to remove, and since the `Photutils` function can only model one galaxy at a time, we first model and remove that BCG, which spans roughly 1000 pixels (0″.06 pixel$^{-1}$). We continue with this logic for the 11 other contaminating galaxies projected near the arc (see Figure 4) and model and remove them according to their projected spatial extent. Figure 5 shows an RGB image created from the original data in the left panel and the cleaned data in the right panel.

Three modeled and removed galaxies presented a problem and may be left out of our data mask, though the reasons why they pose an issue provide us with useful information. The first two galaxies are the ones at the bend at the top of the arc between the leftmost image and the long, thin part of the arc.

---
[2] https://www.stsci.edu/~dcoe/trilogy/Intro.html

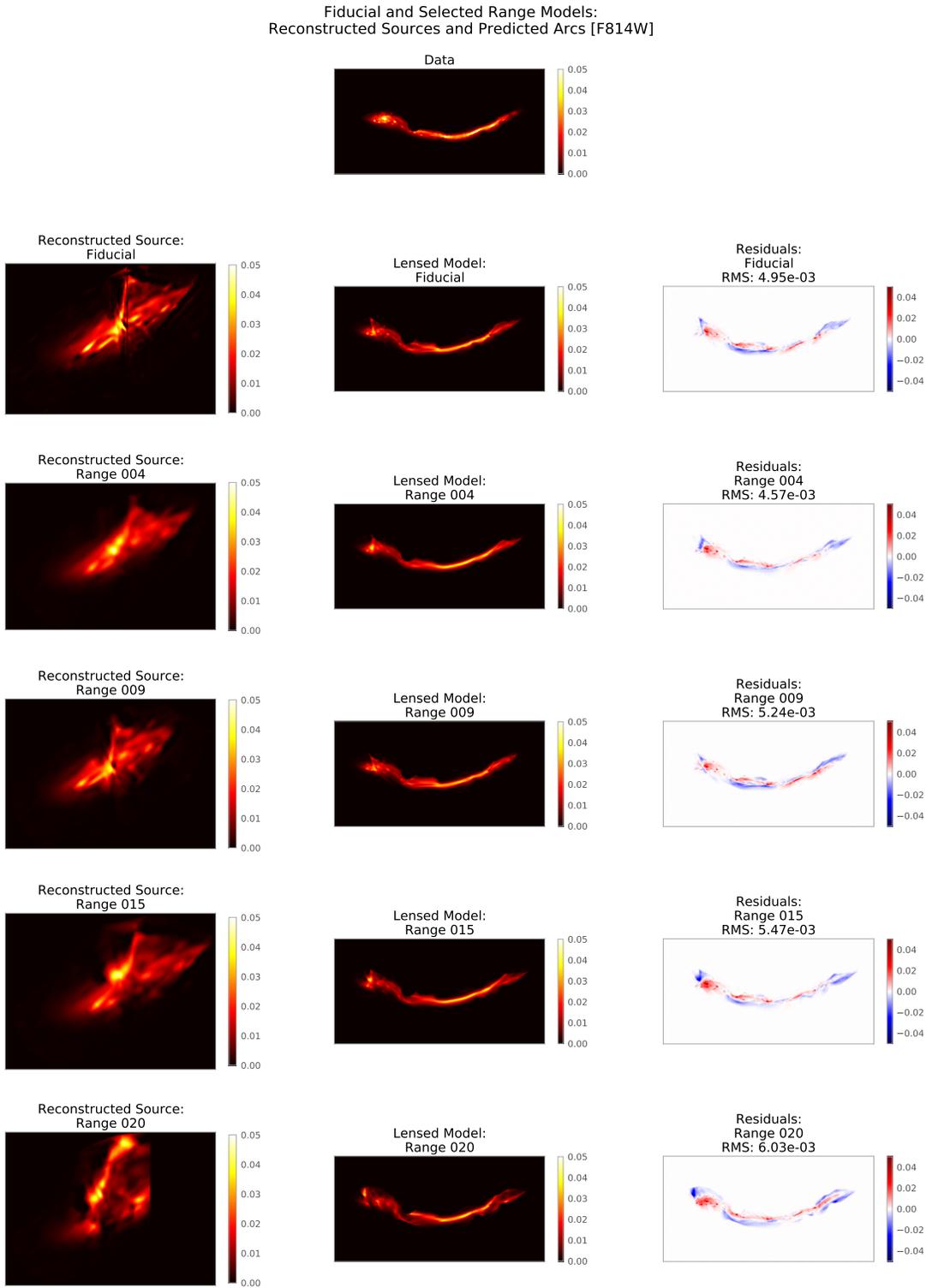

**Figure 7.** Comparison of source reconstructions between the fiducial model and four selected range models ranging from lowest to highest $\chi^2$ values. The first column is the reconstructed source, the second column is the lensed model, and the third column is the residuals between the data and the model. Row 1: original fiducial model; this follows the general shape of the arc and expected source morphology, but false artifacts are visible. Rows 2 and 3: examples of two range models that result in more correctly regularized sources; however, the residuals from the bright, falsely stretched areas near the caustic still result in incorrect arcs and their corresponding patterns in our residuals. Rows 4 and 5: examples of two worse-fitting range models that would be discarded according to our tests; the sources are unphysical and do not resemble any previous source reconstructions, and the lensed models and residuals are similarly ill fitting. Before narrowing down our range of mass models, we need to ensure that the $\chi^2$ and residuals being used for our ranking actually come from the models themselves, not the source reconstruction and regularization methods. All the above tests used an input noise value of 0.015 and a 1/15 resolution grid.

They are similarly small on the sky and projected very close together. Because their light overlaps significantly, we were unable to model and remove them well, so there are circular residuals in the cleaned image. As we will see below, models of the lensed arc need to take care with these two galaxies because they strongly affect the degree to which the arc is bent nearby. (By contrast, treating these two galaxies projected so close as a single mass would not be as problematic for regions farther from the arc.)

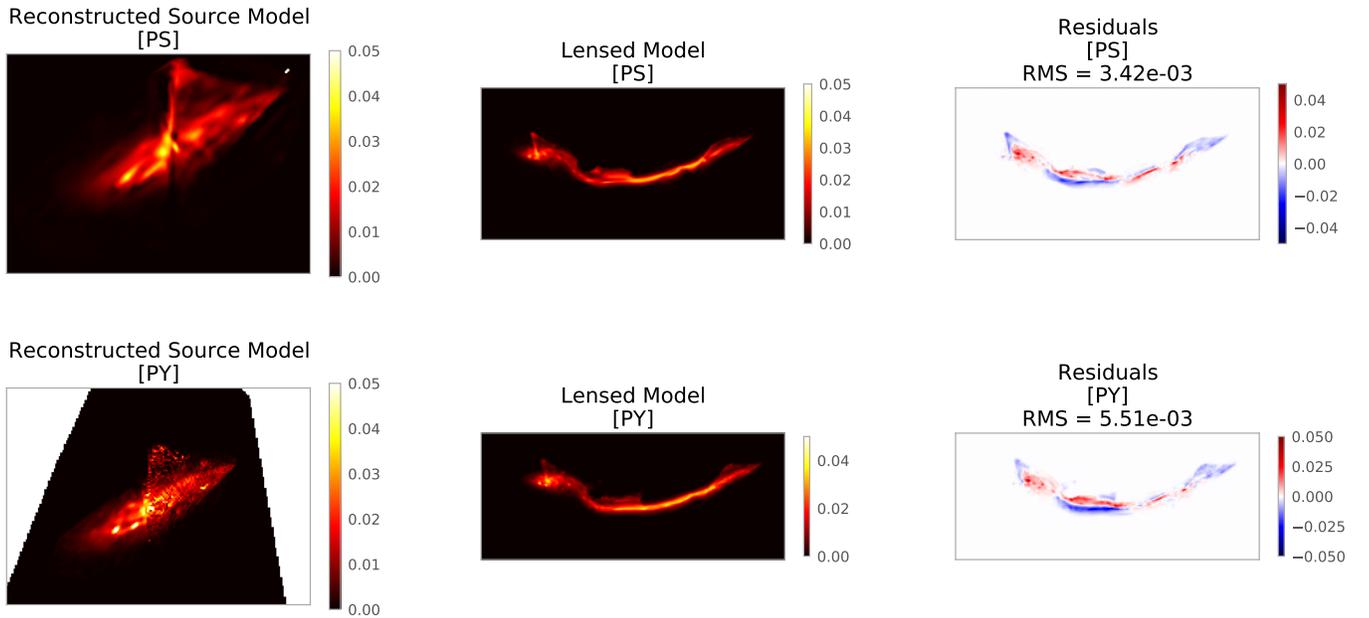

**Figure 8.** Comparison between the `pixsrc` approach and our Python de-lensing code to verify that our methods are similarly reliable for lens model comparisons. `pixsrc` (upper row, noise = 0.015, 1/15 pixels used) + de-lensing code (lower row, full grid used), both run using an edited fiducial model, namely one with equally divided masses and newly estimated positions for the two small galaxies near the bend in the arc. The residual color bars are displayed with the same scale.

The last galaxy we were unable to fit well is the galaxy second from the right on the bottom of the arc. The `Photutils` function we used did not have a PSF input for the galaxy modeling, so some faint diffraction spike-like structure of especially compact parts of galaxies was not caught by the model. This may have caused a part of the arc itself to be oversubtracted if the model extended farther into the arc than necessary. Apart from that, changing different variables in the function such as the amount of sigma clipping or the type of centering consistently left a small residual near the center of the removed galaxy. Though we masked the arc when fitting the galaxies, one reason for this imperfect fit could be the proximity of the galaxy to the arc on the sky, but another reason could be that there might be a faint counter-image, predicted by some of the models we used, lensed by that cluster member.

### 3.4. Data Mask

We used SAOImageDS9 (W. A. Joye & E. Mandel [2003](#)) to take our cleaned image and construct a data mask around the arc in each filter, including a small margin of background around the arc as required by `pixsrc`. We exclude areas with artifacts from residuals, which are more obvious near centers of modeled/removed galaxies, as well a small section of the arc that includes light from the two small/close cluster members that we were unable to remove well near the bend in the arc. The data mask is shown in Figure [6](#).

### 3.5. Noise Level, Regularization, and Grid Size

We estimated the noise level in the data using the standard deviation of pixel values in blank regions of the image. In `pixsrc` and with similar source reconstruction methods, it is often necessary to specify a slightly higher assumed noise value in order to improve the regularization and prevent overfitting (e.g., J. Rivera et al. [2019](#); A. J. Young et al. [2022](#)).

We use two simplifications to reduce the runtime during the initial modeling steps. First, since the HST PSF is quite small relative to the arc, we omit PSF blurring in some steps. Second, for some steps we do not use all of the source pixels when constructing the grid for source reconstruction. We verified using a sample of seven mass models that the ranking of the mass models using their $\chi^2$ values remained consistent with different choices of source grid.

Once the chosen parameters are optimized for a certain series of tests, we then run the `pixsrc` source reconstruction code with full-grid output as well as the PSF. Our preliminary results comparing models using `pixsrc` mainly use the above reduced resolution settings, but our final optimized model results are presented with the full-grid outputs.

### 3.6. Noise Models

In principle, a full Markov Chain Monte Carlo analysis would be the best way to quantify uncertainties in the lens models. In practice, that is too computationally intensive for these models, so we follow the approach from the HFF lens modeling effort and create a set of "noise models" that sample the range of uncertainties. We first add 10 different Gaussian noise maps in Python to create a set of 10 "noisy arcs;" the noise level matches the slightly overestimated noise value from our `pixsrc` inputs that best regularized the data. We then fit our model to each of the noisy arcs, and thus obtain our 10 noise models. We calculate the mean and standard deviation for the two parameters for each galaxy that we varied, and we include these estimates as the model error bars in Section 4.3.1. below.

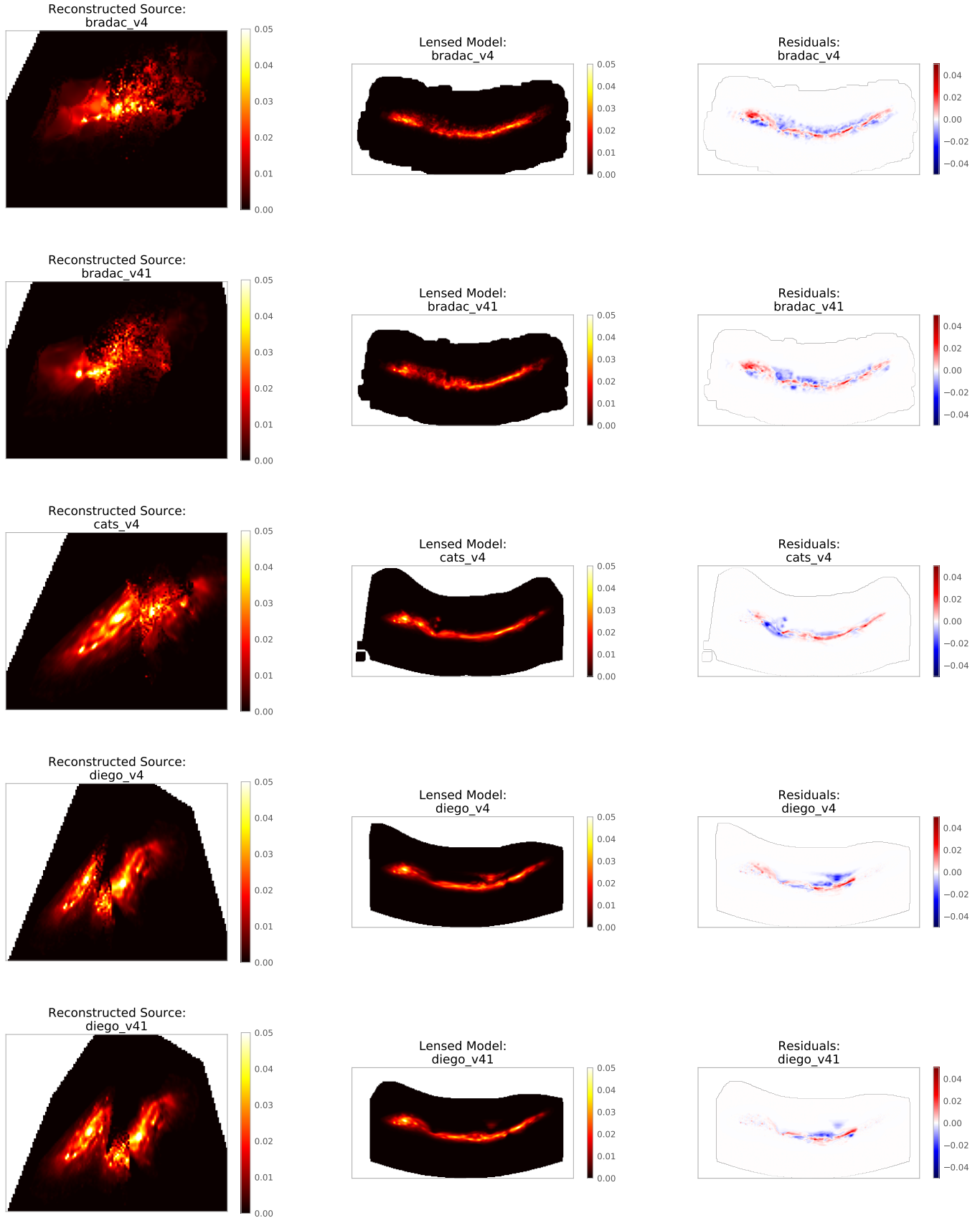

**Figure 9.** Results for five of the public HFF lens models. The left column shows the reconstructions, the middle column shows the model arc, and the right column shows the residuals.

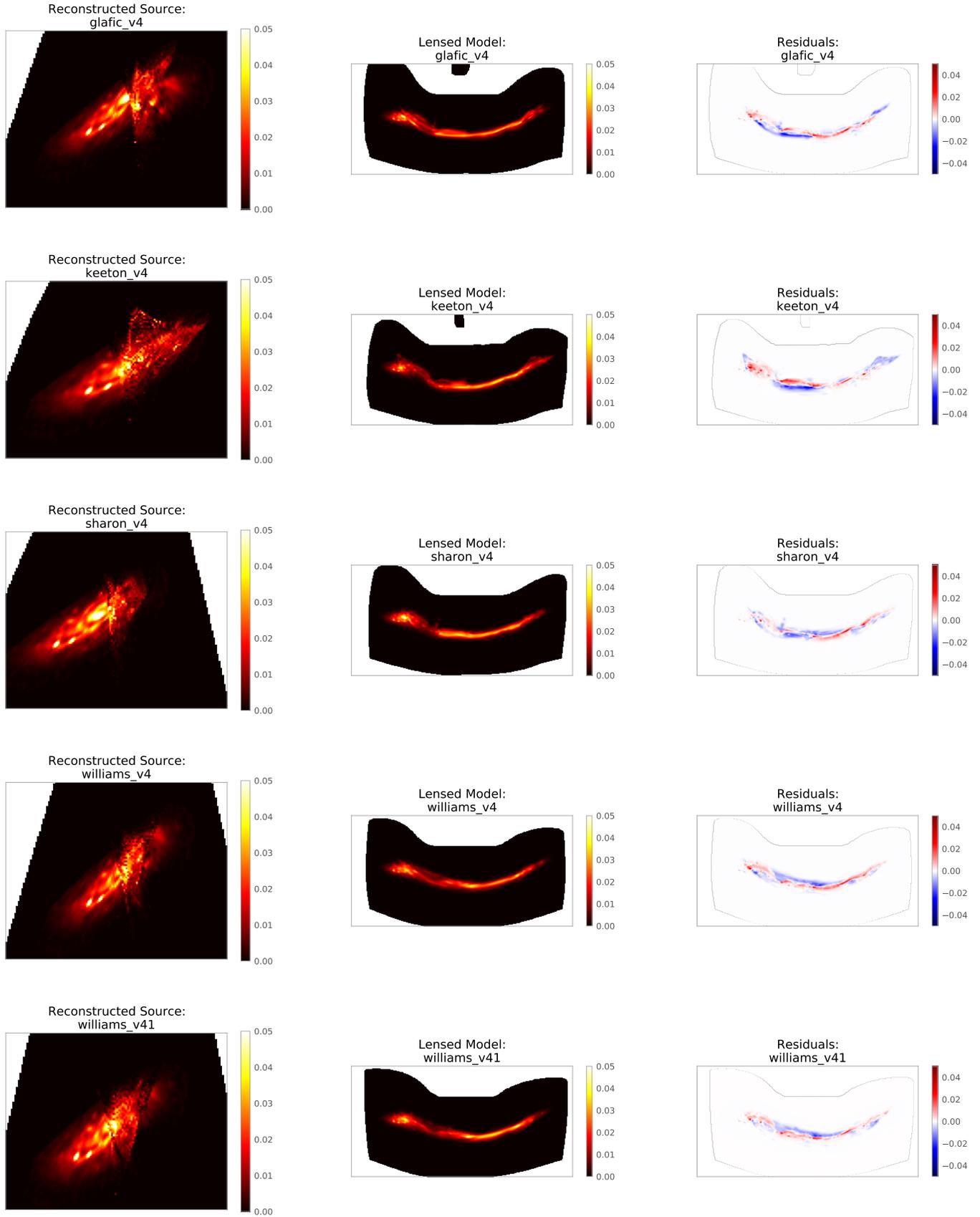

Figure 10. Similar to Figure 9 but for the other five public HFF lens models.

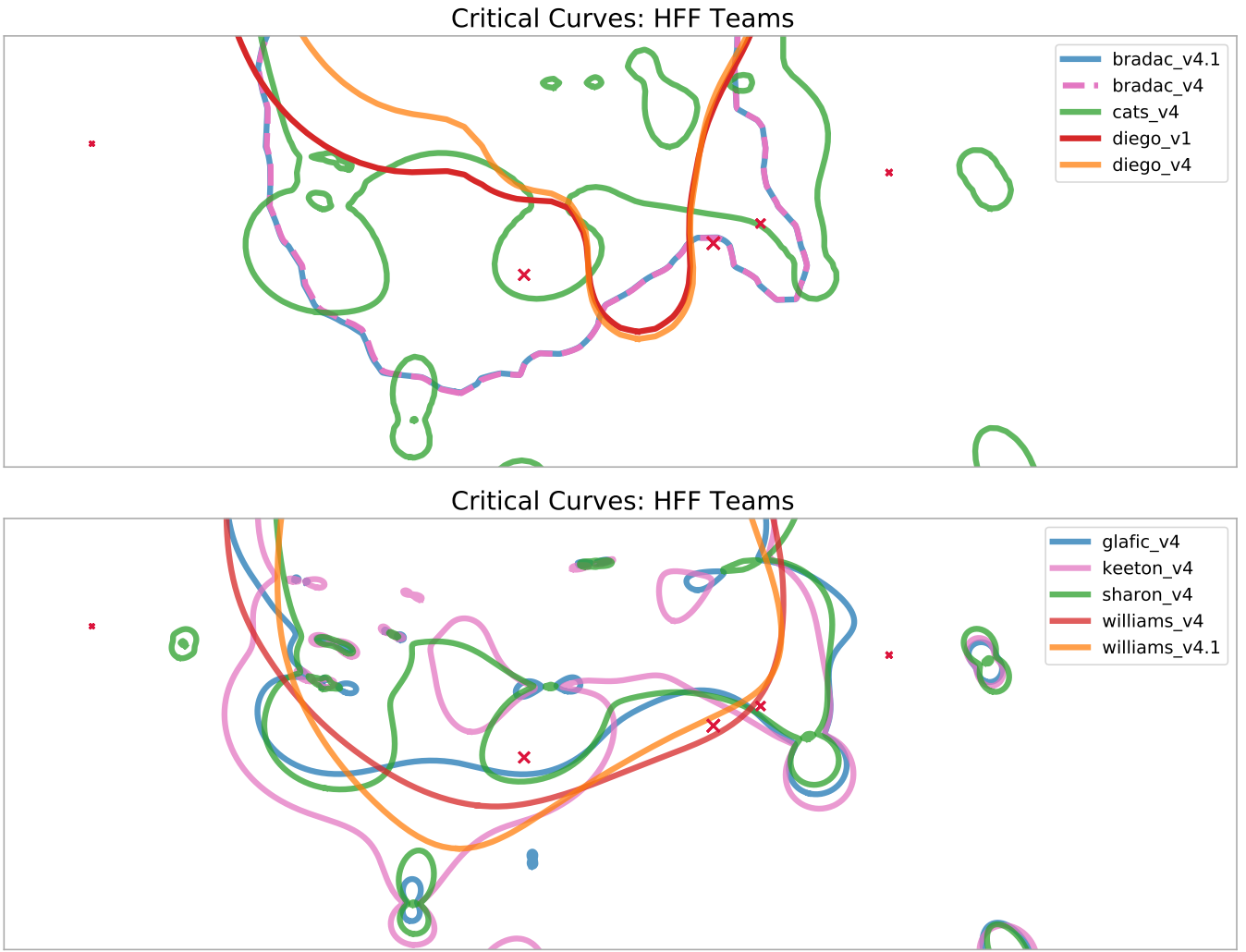

**Figure 11.** Lensing critical curves for the cluster–arc redshift pair, using the lens models from Figures 9 (top) and 10 (bottom). Red × marks indicate the five nominal image centers typically used for lens modeling. Note that the models differ, sometimes significantly, on where the highest magnification regions would be relative to the arc.

## 4. Results

We can now take the masked image containing just the arc light in our three filters and use `pixsrc` (A. S. Tagore & C. R. Keeton 2014) and our new de-lensing code in Python to perform a series of tests, primarily in the F814W filter.

We first use the C. A. Raney et al. (2020a) fiducial and range models to reconstruct the lensed arc and see how well the brightness distribution of the arc is reproduced by that model. The range models sample the range of uncertainties in the fiducial model, which would help us determine whether reconstructing the source using all the extended arc pixels can be used to rule out lens models that would fail to fit them. These existing mass models were constrained using point sources from lensed images; for each `pixsrc` run, we compare the data image to the model arc, which allows us to understand (a) how well each model reconstructs the rest of the arc pixels and (b) where and why the model fails.

We then compare the C. A. Raney et al. (2020a) fiducial and range models to the fiducial and available range models from the other HFF lens modeling teams.

Finally, we take the C. A. Raney et al. (2020a) fiducial model, correct the inputs, and vary the parameters using `pixsrc` to improve our lens model with respect to the giant arc brightness distribution.

### 4.1. Fiducial and Range Models: Initial

We begin with the fiducial mass model and range mass models from C. A. Raney et al. (2020a). Figure 7 shows our comparison between the fiducial, best-fit range, and worst-fit range models. The fiducial model is able to fit the overall shape of the arc but deviates in certain regions, notably around the bend in the arc. The range models differ in some details but likewise cannot accurately reproduce the arc; furthermore, some of the models yield sources whose structures are implausible (notably the bottom two rows). We can thus see that our current range of models cannot describe the arc and this high-magnification region correctly, and adjustments need to be made in order to do any detailed studies of the arc as a whole in those regions.

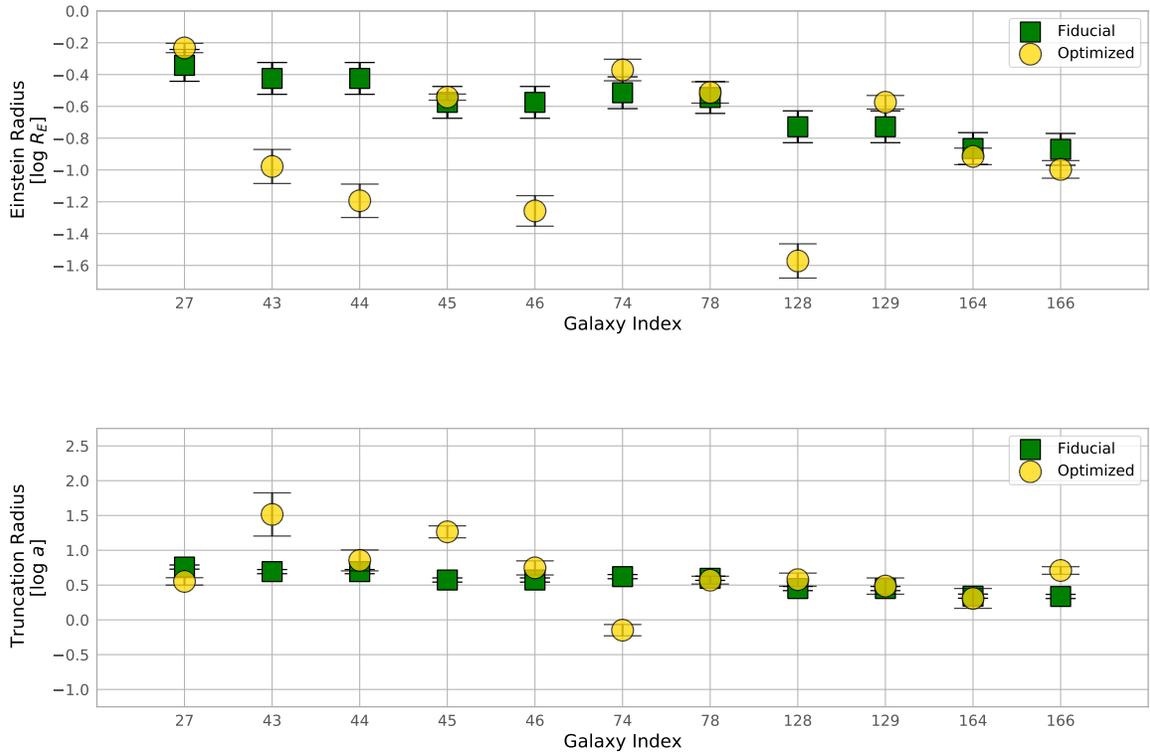

**Figure 12.** Here we show the changes in parameter values between the old fiducial (green squares) and optimized (yellow circles) mass models. The error bars on the fiducial model reflect scatter in the mass–luminosity scaling relations (see Section 2.2), while the error bars on the optimized model come from our noise model analysis (see Section 3.6). The top panel shows the Einstein radius in arcseconds, scaled to the cluster–source redshift pair; the bottom panel shows the truncation radius in arcseconds. The galaxy indices are the same as in Figure 4. The galaxies that changed the most are the ones near the bend in the arc on the left where the fiducial model failed the most noticeably in Figure 8, as well as in the region where the fiducial model missed the arc on the rightmost side.

Before turning to lens models from other teams, we test the source reconstruction from `pixsrc` against the source reconstruction from our new Python code (Section 2.4). Figure 8 shows that the two codes yield similar results for the reconstructed sources, model arcs, and residuals. While the rms values of the residuals differ quantitatively, the structure of the residuals is similar between the two methods. In additional tests, we found that `pixsrc` and our Python code give consistent results for ranking models by fit quality. We conclude that our new Python code offers a reliable way to evaluate models for this arc.

### 4.2. Hubble Frontier Fields Fiducial Models: Comparison

We now turn to models from the other HFF lens modeling teams (listed in Section 2.2). Figures 9 and 10 show the results for the fiducial lens models from each team. None of the models constrained using various parametric, nonparametric, and hybrid methods, including using point-like image positions, were able to de-lens to a reasonable source (whose morphology should be similar to the less complex source reconstruction using only the leftmost image of the arc in J. Richard et al. 2010; V. Patrício et al. 2018). Some models lack the structure that would match the spiral morphology seen in the leftmost image of the arc; some de-lens the galaxy into two false, spatially separate, twin sources; some of the models fail in similar ways as the fiducial one from C. A. Raney et al. (2020a). Using the same process, we tested the publicly available range models that varied the properties of each cluster model; they were all similar to their teams' fiducial models in the ways they were not able to fit the arc well, implying that the mass distribution of the cluster has not been well understood beyond the expected levels of uncertainty.

In addition to examining the reconstructed sources and model arcs, we plot the lensing critical curves using the publicly available convergence and shear maps scaled to the cluster–source redshift pair in Figure 11. The variation among the models in these high-magnification regions helps us understand the corresponding variation among the source reconstructions and the location of the more significant structure in the residuals from the extended giant arc. Better constraining these regions would lead to more accurate predictions about the magnifications, time delays, and properties of the source.

### 4.3. Optimized Model

Since existing lens models do not fit the arc well, we return to the fiducial model from C. A. Raney et al. (2020a) to see if we can identify the problems and then improve the model. One of the major deviations from the data occurs in the bend in the arc to the right of the leftmost image of the background galaxy. As discussed in Section 3.3, there are two small galaxies projected close to each other and resolved in the HFF data, but they were treated previously as a single mass. As a first step, we separated these galaxies, estimated their $x$- and $y$-positions, and assigned each one half of the original luminosity for the purpose of estimating their Einstein radius and truncation radius. The number of galaxies in our lens model thus increased by one to 276.

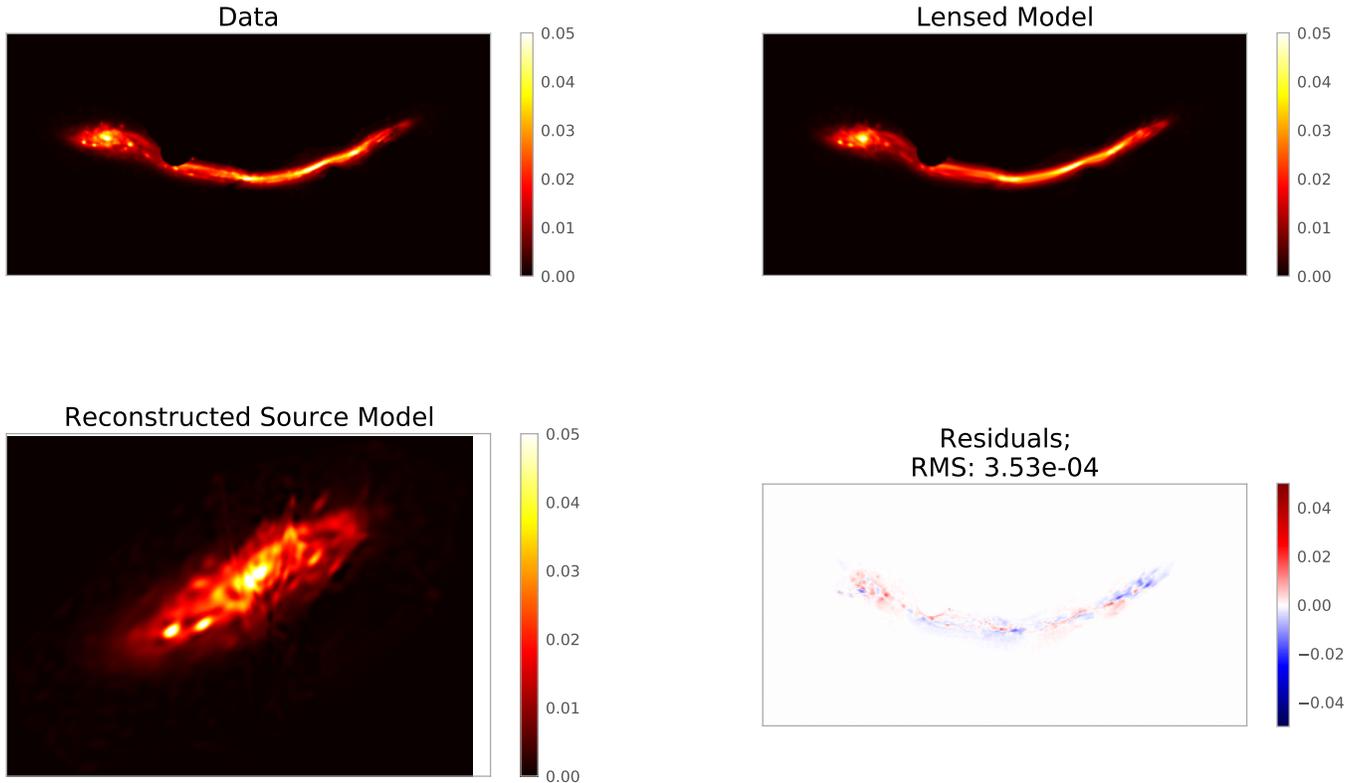

**Figure 13.** Full-resolution `pixsrc` output for the F814W filter arc data using the optimized lens model after separating the two small galaxies near the arc and varying the lensing strength and truncation radius of 11 lensing galaxies local to the arc.

We then optimized the model by allowing `pixsrc` to vary the Einstein radii and truncation radii for the 11 galaxies near the arc that are shown in Figure 4. We continued to impose priors using the luminosity–mass scaling relations, with a scatter of 0.1 dex for the Einstein radius and 0.03 dex for the truncation radius.

*4.3.1. Parameter Changes*

We report the changes in parameter values between the fiducial and optimized models. The lensing strength has a mass–luminosity ($M/L$) scaling relation scatter of 0.1 dex from the C. A. Raney et al. (2020a) mass model; the truncation radius which is the truncation radius, similarly has an associated $M/L$ scatter of 0.03 dex. We visualize the changes in these two parameters in Figure 12, in which deviations from the individual points represent the parameters of the masses that were varied and optimized. We found that galaxies 45, 46, 43, and 128 (see Figure 4) had the most drastic changes compared to their original values in the fiducial model. Our iterative process in determining how to optimize which parameters for which galaxies illustrates how source reconstruction can be useful for identifying issues or incorrect assumptions in a lens model and correcting them by location and expected mass.

In the HFF data, the galaxies we labeled 45 and 46 were somewhat resolved to be separate, but most of the mass models from HFF teams used a single point for them from previous observations; in high-resolution JWST data from Y. Fudamoto et al. (2025), these two galaxies are even better resolved and separated. We recognize that galaxies 45 and 46 are the two galaxies closest to the bend in the arc where the fiducial model was the most unable to accurately model the arc; since we first spatially separated them into two different inputs, our optimization tests further refined their best-fit parameters. Galaxy 43 has a light distribution in HFF data more consistent with a possible, less massive spiral galaxy rather than a massive elliptical galaxy, but it was assumed to be an elliptical cluster member in the fiducial model we used; we note that the parameter representing the lensing strength (proportional to the mass) of this galaxy was determined to be much less than what was originally assumed in the fiducial model. New JWST data from Y. Fudamoto et al. (2025) show this morphology even more clearly, which is likely a spiral galaxy. Galaxy 128 is projected close to the rightmost region of the arc where the fiducial model arc deviated from the data; its light may have been contaminated by arc light, which may have lead to an incorrect mass estimation from the $M/L$ scaling relation input to the fiducial lens model. In all four areas, the results of the optimized source reconstruction are much more aligned with the giant arc data, which is evident in the residuals.

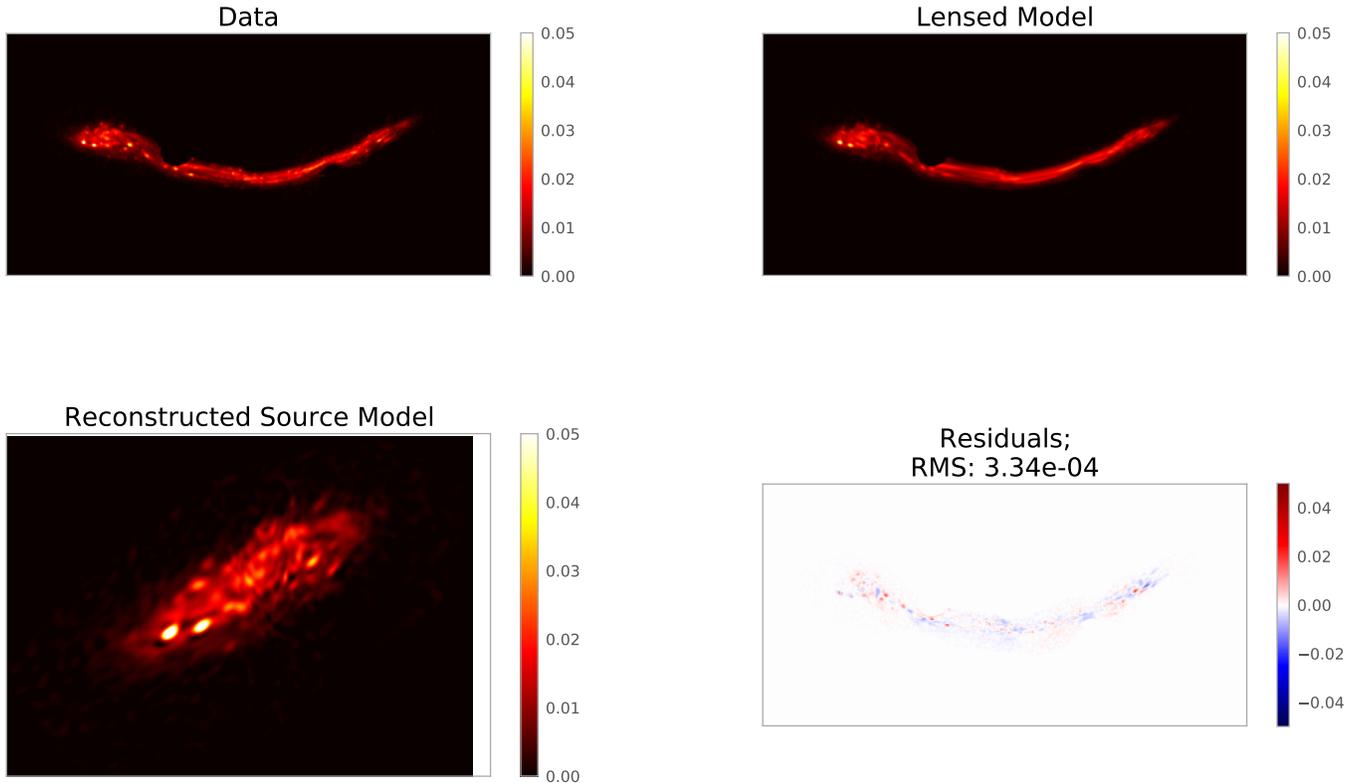

**Figure 14.** Same as Figure 13 but for the F606W filter arc data.

The issues revealed and at least partially corrected using PBSR of giant lensed arcs to locally constrain parts of a cluster mass model could potentially be useful in studies where lower-resolution data might not lead to a good enough mass model on its own; if an extended giant arc is present, better constraints can be put on the data until a higher-resolution follow-up study can be done or in lieu of one if the cluster is a difficult target.

### 4.3.2. Full Pixel-based Source Reconstruction

We now run `pixsrc` with the full source grid output and the PSF. The final, optimized lens model in Figure 13 is better able to reproduce the input arc, as seen in both the improved residuals (especially near the aforementioned bend in the arc) as well as the lower rms value. We present the full-grid, PBSR in three HST filters in Figures 13–15. Note that in each of these source reconstructions and predicted models, the unphysical effects along the path of the caustics and critical curves have been corrected, and the spiral structure of the galaxy better matches what is visible in the least-warped image in the data.

The optimized model created with the F814W filter fits the general shape and the finer detail of the arc well for both the F814W (red) and the F606W (green) filters. It matches the overall shape of the arc well for the F435W (blue) filter, but, perhaps because of a different noise level in the data, it does not fit some of the finer details, namely the star-forming clumps. These regions include what was used in studies like J. Richard et al. (2010) and V. Patrício et al. (2018) to constrain their mass model, but those models did not include the entire brightness distribution of the arc. Ideally, all three filter source reconstructions could be utilized as constraints in an optimized mass model; we will incorporate this into our future Python code and attempt to include all possible constraints from the extended brightness distribution in all three filters.

### 4.3.3. Critical Curves, Caustics, and Image Positions

Since we only varied galaxies near the arc, we expect major changes to the lens model to be concentrated in this region. One way to visualize the changes is to compare the fiducial and optimized lensing critical curves. Figure 16 shows that there are indeed critical curve variations near the arc that are needed in order to fit the arc better. Away from this region, the old and new critical curves are very similar.

Another way to examine this effect is to apply the optimized model to the point-image data that were originally used to construct the fiducial model. C. A. Raney et al. (2020a) found rms residuals between the observed and model positions of $0\rlap{.}''71$ for the fiducial model. Our optimized model yields rms residuals of $0\rlap{.}''75$. In other words, the local changes in our optimized model do not significantly affect the bulk of the HFF images that are distributed over a much larger field.

### 4.3.4. Derived Source Properties

In Figure 17, we show a full-color RGB representation using the source reconstructions from the three HFF filters. We visually fit an ellipse to the spiral galaxy and note that the semimajor axis is ~12.63 kpc in the source plane; the bluer star-forming regions visible on the bottom left section of the galaxy are on the order of 1 kpc.

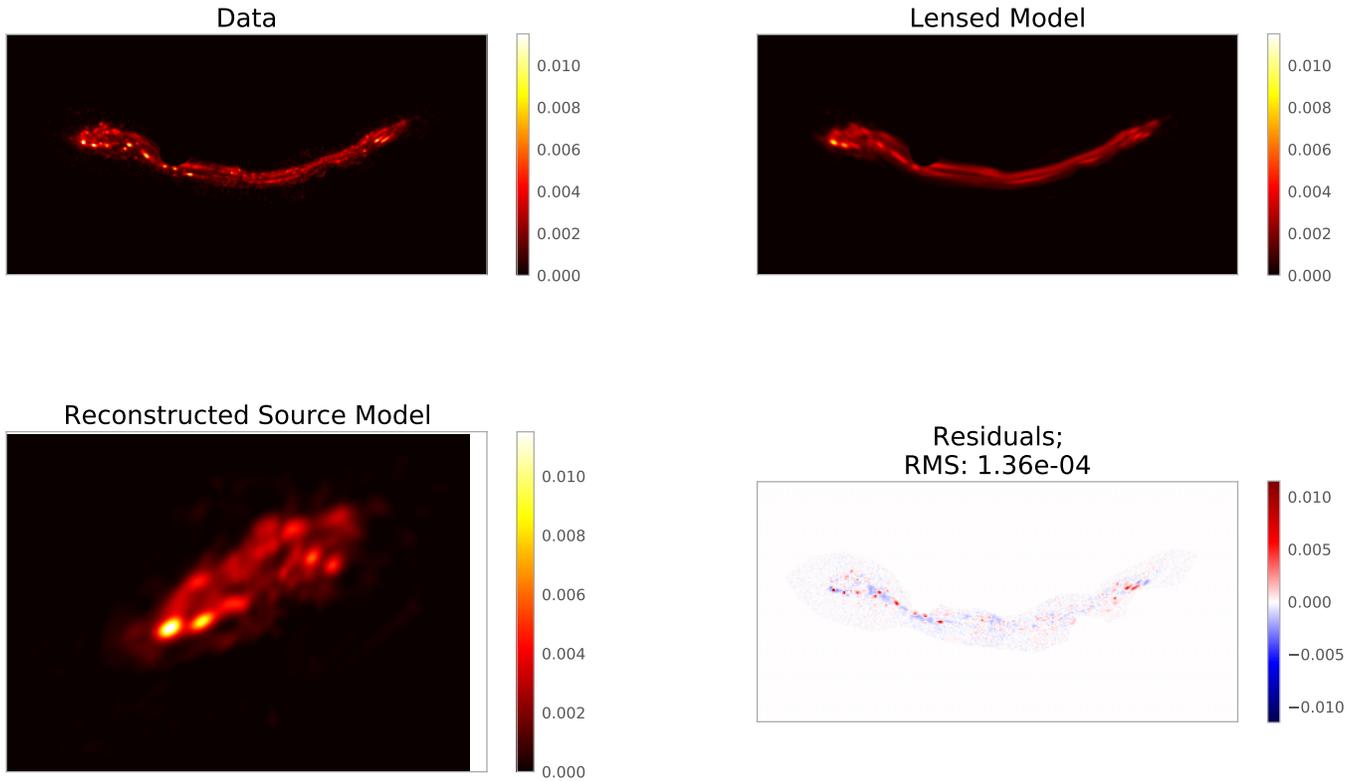

**Figure 15.** Same as Figure 13 but for the F435W filter arc data.

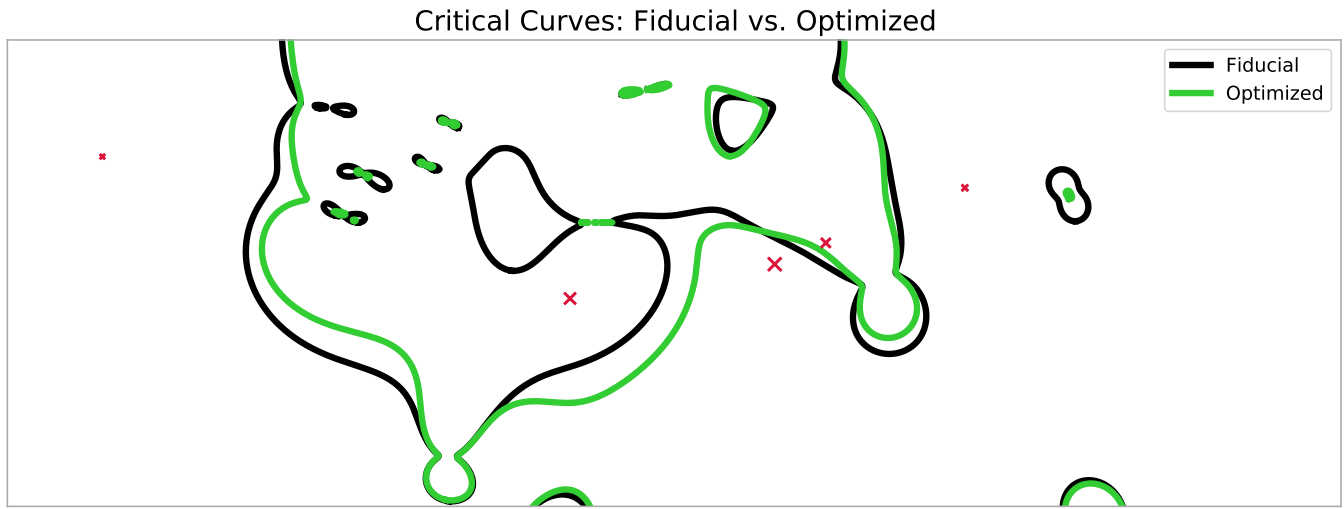

**Figure 16.** Lensing critical curves for the old fiducial model (black) and the newly optimized model (green).

We show the critical curves overlaid on the model image as well as the caustics overlaid on the source reconstruction. The RGB image allows us to note that the positioning of the redder source galaxy center relative to the different caustics it crosses in the source plane indicates that it is this central region that gets stretched out in the most highly magnified regions of the arc that are visible in the image plane giant arc.

The complex structure of the galaxy cluster makes the caustics more complicated than what would be seen from a single lens, which is especially noticeable for regions producing several lensed images (C. R. Keeton et al. 2000). There are tangential (J. Wambsganss 1998) caustics from the larger-scale cluster mass components, which can be seen in the large diamond-shaped yellow lines. There are also smaller-scale caustics that originate from the figure-eight-shaped loops of the critical curves of smaller galaxies embedded within the larger-scale critical curves in the bottom left panel.

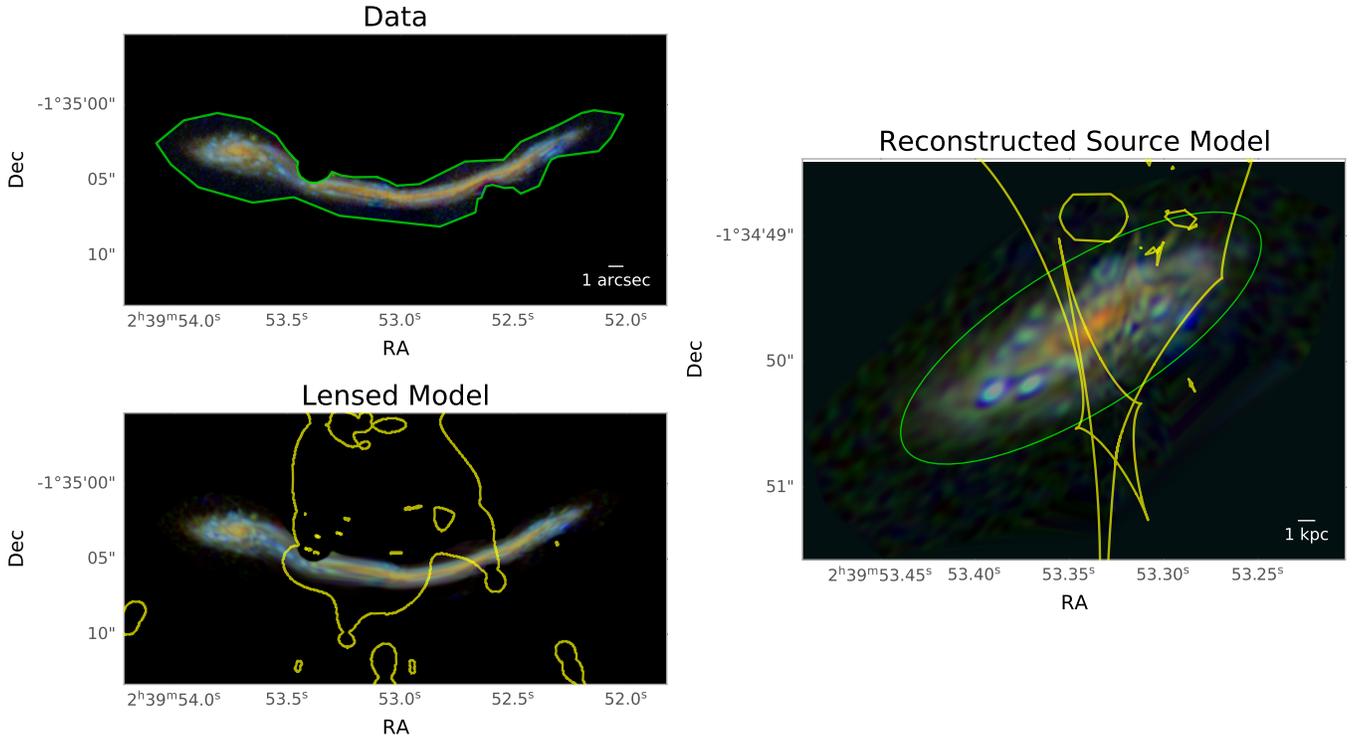

**Figure 17.** RGB images constructed using the F814W, F606W, and F435W results in Figures 13–15. The top left panel shows the data, with a 10 kpc scale bar indicated. The bottom left panel shows the lensed model, with the lensing critical curves for the cluster–source redshift pair shown in yellow. The right panel shows the reconstructed source, with the lensing caustics shown in yellow. The green ellipse represents a visual estimate for the size and shape of the source: it has a semimajor axis of 12.63 kpc. A 1 kpc scale bar is shown for reference for the physical size of the star-forming regions and other features in the source plane.

## 5. Conclusions

We have demonstrated that the current method for constraining cluster models is insufficient for detailed studies of individual giant arcs. We have also confirmed that our updated model, which better reproduces the arc, only required changes in galaxy masses projected as local to the arc, as opposed to global masses on the scale of cluster dark matter halos and X-ray gas. This can be used to correct any galaxy-scale assumptions made at the start of the analysis before they propagate into derived source property calculations or galaxy cluster implications derived from the lens modeling.

Our results illustrate how fine-tuning cluster models using giant arcs informs us about the smaller-scale, local mass distribution in the cluster, but the global cluster mass assumptions may remain unchanged. This shows us that it is critical to test and refine a cluster mass model using any available extended giant arcs before doing detailed studies of the arc. It is not enough to simply constrain the cluster model using more point-like image positions of different source features in the case of less uniform sources. The more warped, higher-magnification regions in a giant arc hold information that is key to constrain, and testing the model with just one of the merging extended images or with separate extended images may not be able to match the same information about the position and brightness of each lensed pixel from different merging images in the arc.

Our study using a particularly resolved giant arc from the HFF shows how it is important to make use of the full information given to us by the fortuitous alignment that leads to giant arcs lensed by galaxy clusters, as there are still inconsistencies between models and data and between each of the models themselves, even with a wide variety of lens modeling techniques representing the state of the field.

This has implications for the conclusions drawn from both giant arc studies reliant on source reconstruction as well as galaxy cluster studies in general. Without fully utilizing the data afforded by giant arcs, we may end up with incorrect inferences about both the cluster mass distribution that leads to the details of lensing of the extended brightness distribution as well as the higher-redshift source itself. In the case of A370, the more detailed study of the giant arc led us to correct certain assumptions made from measurements using lower-resolution data, namely the formerly unresolved cluster members near the bend in the arc to the right of the leftmost arc image and the classification of the morphology of a local galaxy. For source studies, values calculated from the data such as star formation rate, kinematics, and the times or locations of transient events could be affected.

Studies such as J. M. Diego et al. (2024), Y. Fudamoto et al. (2025), and M. Limousin et al. (2025), who noted that some of their JWST observations of microlensed stars in the A370 giant arc were farther away from the critical curves of any available cluster model for this field than expected, could potentially draw incorrect conclusions about the reason for the disagreement and attribute incorrect underlying reasons for it. Potential reasons given were dark matter substructure, which could be studied more in-depth if systematic uncertainties in the mass model were small enough, or a shallow stellar luminosity function, whose distribution could be misinterpreted with the small magnified sample available with lensing. High-redshift studies of magnified objects, especially with new, high-resolution data, are highly dependent on understanding the uncertainties inherent to lensing. Finding ways to efficiently reduce the systematic uncertainties in any cluster modeling codes and methods, which C. A. Raney et al. (2020b) found to be larger than any remaining statistical uncertainties, is key to ensuring we do not make incorrect assumptions about the high-redshift Universe and its evolution throughout history.

In other ways, being able to streamline constraining galaxy cluster lenses using giant arcs could lead to a better understanding of dark matter distribution in galaxy clusters at different redshifts, which could be useful in constructing machine learning training sets to then be used on (1) regions in the same galaxy cluster without giant arcs or (2) other galaxy clusters completely without giant arcs. This could be used to add a layer of finer-scale cluster mass distribution corrections to surveys such as the established Sloan Digital Sky Survey Giant Arcs Survey (K. Sharon et al. 2020). It could also help determine follow-up targets with interesting cluster mass distributions from lower-resolution, large-scale surveys such as the upcoming Legacy Survey of Space and Time (LSST Science Collaboration et al. 2009).

Our next step is to further develop and expand our Python source reconstruction code so that it can be used with the new Python version of lensmodel, pygravlens (C. Keeton 2025) and use it to establish a straightforward, standard methodology for modeling galaxy clusters with caustic-crossing giant arcs in the field.

## Acknowledgments

We thank Catie Raney, Anthony Young, and Somayeh Khakpash for helpful discussions. We thank the anonymous referee for helpful suggestions. We acknowledge financial support from the US National Science Foundation through grant AST-1909217 and from the Rutgers—New Brunswick Department of Physics & Astronomy through a Claud Lovelace Graduate Fellowship (L.E.).

## Data Availability

The data used in this paper were published as part of the HFFprogram; see J. Lotz (2013).

## ORCID iDs

Lana Eid  https://orcid.org/0009-0003-5696-9355
Charles R. Keeton  https://orcid.org/0000-0001-6812-2467